\documentclass[aps,prl,groupedaddress,showkeys]{revtex4-1}

\usepackage{graphicx}
\usepackage{bm}
\usepackage{amssymb,amsfonts}
\usepackage{amsmath}
\usepackage{graphicx}
\usepackage[usenames,dvipsnames]{xcolor}
\graphicspath{{Figures/}}
\usepackage{soul}

\begin{document}

\title{Spacetime Foam and Solution of the Cosmological Constant Problem }


\author{Alexander I Nesterov}
  \email{nesterov@academicos.udg.mx}
\affiliation{Departamento de F{\'\i}sica, CUCEI, Universidad de Guadalajara,
Av. Revoluci\'on 1500, Guadalajara, CP 44430, Jalisco, M\'exico}

\date{\today}

\begin{abstract}
The cosmological constant problem is a fundamental issue that has puzzled researchers in the fields of theoretical physics and cosmology for a long time. It arises from the discrepancy between the observed value of the cosmological constant and the value predicted by quantum field theory. A new spacetime model based on nonassociative geometry and statistical physics of complex networks offers a fresh perspective on the problem. Our research indicates that spacetime topology plays a crucial role in solving the cosmological constant problem and addressing the dark energy issue. We discovered that spacetime foam significantly impacts the effective cosmological constant, which is determined by the density of topological geons. Furthermore, we demonstrate that the source of dark energy is topological geons
\end{abstract}




\maketitle

The equivalence principle of general relativity requires that every form of energy gravitate similarly. According to this principle, the enormous energy of the quantum vacuum fluctuations must produce a sizeable gravitational effect. In particular, it implies that the virtual particles' contribution to the vacuum energy density, commonly referred to as the zero-point energy density, should affect the value of the cosmological constant \cite{ZEYB,ZEYB1}. Quantum field theory (QFT) allows us to calculate the contribution of the zero-point energy to the cosmological constant. Unfortunately, the estimates disagree with observational data by $1 0 ^{122}$ factor. This discrepancy between theory and observation is a cosmological constant problem. It is also known as the worst prediction of theoretical physics \cite{WEST1,CSPWTE,DAD,CASM,PPRB,PADM1,CESMT,RUGH,ORCOM,SVSA,PPJRB,WEST}. 

Despite extensive research, the cosmological constant problem remains an open and unsolved puzzle in physics. Some proposed solutions to this problem invoke the anthropic principle, suggesting that we live in a universe where the cosmological constant is small enough to allow for the existence of intelligent life \cite{WEST,WEINST,CBJRM,WF1,BJTFW,BECA}. Others involve modifications to quantum field theory or general relativity, including emergent gravity, that implies omitting the use of continuum concepts {\em a priori}. It is assumed that at the Planck scale, spacetime is fundamentally discrete 
\cite{BI,MJPJ,DO3,Loll-1998,tHGer1,VRMA,SWBR,PATH1,DOFAY,PATH2}.

In the middle of the last century, Pauli suggested that the vacuum energies of bosons and fermions might compensate for each other \cite{WPECW}. This assumption is based on the fact that the vacuum energy of fermions and bosons have opposite signs: the energy of bosons has a positive sign, while that of fermions has a negative one. Later, Zel'dovich developed this approach to link the vacuum energy to the cosmological constant \cite{ZEYB,ZEYB1}. 

Using the Pauli-Zel'dovich idea of cancellation of vacuum bosonic and fermionic degrees of freedom, we have recently shown that the ``foamy" structure, analogous to Wheeler's ``spacetime foam" \cite{WHJA1,WHJA}, significantly contributes to the effective cosmological constant. The latter is defined by the universe's Euler characteristic \cite{NAHM}. In this Letter, we describe this approach in detail and explain the contribution of the topology and curvature to the cosmological constant. We show that the density of topological geons determines the cosmological constant. 

We use the natural units,  setting $ c= \hbar =k_B =1$.
\\
 
{\em Building discrete spacetime.} --  Our approach is based on nonassociative geometry and statistical physics of complex networks and suggests a unified algebraic description of both continuum and discrete spacetime as well \cite{NS3a,Sab1,NS3,Sab2001,NS4,N2006,NAMH,NAHM}. We treat space as a complex network presented by simplicial 3-complex. Nodes of the network (0-simplices) are the atoms of spacetime (pre-quarks or preons) that represent the universe's building blocks, just as atoms are the building blocks of matter. They are assumed to be fermions with spin one-half and size $\sim \ell_p$. An edge, or 1-simplex, has two vertices; therefore, it is a  boson composed of two preons. A 2-simplex being built from three preons is a fermion. A 3-simplex is a boson formed by four preons.

 The curvature of the simplicial complex is described by an elementary holonomy -- non-local analog of curvature \cite{Sab1,NS3a,NS3,Sab2001,NS4,N2006,NAMH}. The elementary holonomy is associated with the faces of  2-simplex and can be written as $h= 1 \pm\Delta S/R^2$, where $\Delta S$ is its area, and $1/R^2$ is the curvature. Upper/lower sign corresponds to positive/negative curvature, respectively \cite{NAMV,NAHM}.  
  
\begin{figure}[tbh]
 \begin{center}
  \includegraphics[width=3.5cm]{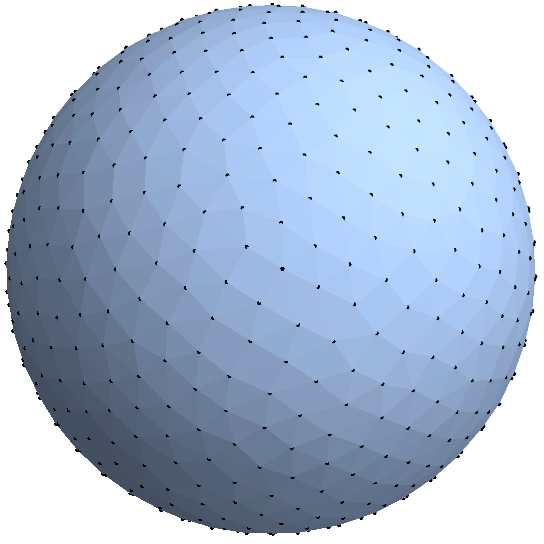}  \quad
  \includegraphics[width=3.5cm]{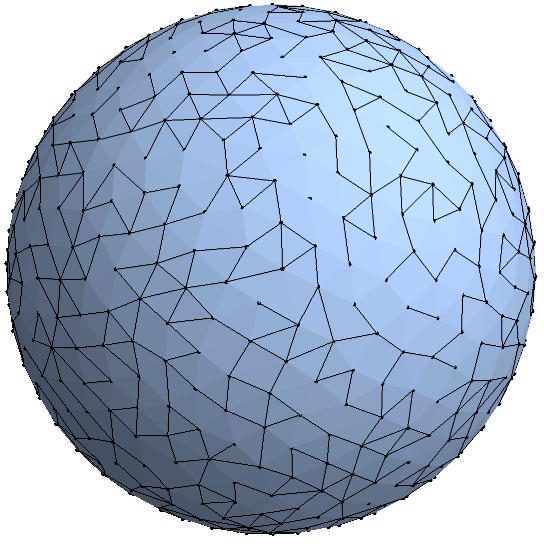}\\ 
  \includegraphics[width=3.5cm]{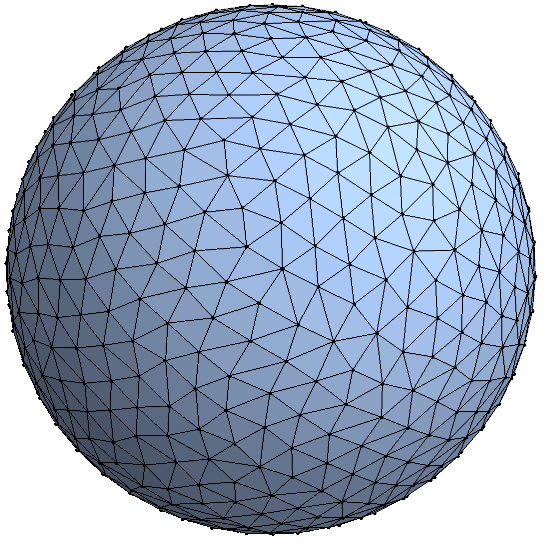}
  \end{center}
  \caption{Discrete space with positive curvature.  Left top: a simplicial 0-complex (a non-triangulated, completely disconnected space). Right top: a simplicial 2-complex (a partially triangulated space). Together with isolated 0-simplices and 1-simplices, one can observe the formation of two-dimensional clusters connected by simplicial 1-complexes. Bottom: a pure simplicial 2-complex (a triangulated space). Adapted with permission from Ref. \cite{NAMH}. }
  \label{fig1}
\end{figure}

  A finite simplicial complex is pure if each facet has the same dimension. In other words, a pure simplicial complex is formed only by simplices of the same dimension. A pure simplicial complex with slight variations of the edge length can be regarded as a discrete approximation to a smooth manifold. However, some simplices can collapse to lower dimension simplices. In this case, the simplicial complex can not be considered as a decomposition of a smooth manifold \cite{HAWK1}. As an example, consider the simplicial network with positive curvature presented in Fig. \ref{fig1}. One can observe a non-triangulated, completely disconnected space (top left), a partially triangulated space (top right), and a pure simplicial 2-complex (bottom) that can be thought of as a triangulation of the two-dimensional sphere.

{\em Statistical description of complex networks.} --  The statistical description of complex networks is based on the Shannon-Gibbs entropy,
$S = - \sum_{G \in \mathcal G} P(G) \ln P(G)$, where  $P(G)$ is the probability of obtaining a graph $G $, belonging to ensemble of graphs $\mathcal G$ \cite{PJNM,STCD,HPLG}. For an undirected network with a fixed number of vertices and a varying number of links, $P(G)$  is given by \cite{PJNM,CDLM,CDAS,CGST}
\begin{align}
P(G)=\frac{1}{\mathcal{Z}} \exp \big (\beta(\mu L(G)-H(G))\big), 
\end{align}
where $\mathcal{Z} =\sum_{G \in \mathcal G}  \exp \big (\beta(\mu L(G)-H(G))\big),$ is the partition function,  $\beta =1/T$ is an inverse network temperature, $\mu$ is the chemical potential, and $L(G)$ is number of links in the graph $G$. 

Let us assign the ``energy", $\varepsilon_{ij}$, to each edge $\langle i,j \rangle$. Then the graph Hamiltonian can be written as $H(G) =\sum_{i<j} \varepsilon_{i j} a_{i j}$, 
where $a_{ij}$ is an adjacency matrix. It takes value 0 or 1 in the $i,j$ entry for each existing or non-existing link between pairs of nodes ($ij$). The computation of the
partition function yields \cite{CDAS}
\begin{align} \label{Z}
    \mathcal{Z}=&\sum_{G \in \mathcal G} \prod_{i<j} e^{\beta\left(\mu-\varepsilon_{i j}\right)  a_{i j} }.
\end{align}

The network connectivity is characterized by the connection probability $p_{ij}$, i.e., the probability that a pair of nodes $(ij)$ is connected. Using the partition function \eqref{Z}, one can obtain the connection probability of the link between nodes $i$ and $j$ as the derivative of the partition function: $p_{i j} =- {\partial \ln \mathcal{Z}}/{\partial (\beta \varepsilon_{ij})}$ \cite{WDSS,NMSW,BSLV,ARB}.

For a {\em fermionic graph}, when only one edge is allowed between any pair of vertices, we obtain the Fermi-Dirac distribution \cite{PJNM}:
\begin{align}
    p_{i j}=\frac{1}{e^{\beta \left(\varepsilon_{i j}-\mu\right) } +1},
    \label{FDBE}
\end{align}


{\em Model.} -- In our concept, spacetime comprises entangled quanta (preons), similar to how quantum spin liquids arise from the collective behavior of entangled spins. Gravity and geometry are emergent phenomena occurring in the low-energy corner of the physical fermionic vacuum. At high temperatures, $T \gg T_p$, where $T_p$ is the Planck temperature, spacetime behaves like a quantum Fermi liquid. At the Planck temperature, the topological phase transition occurs, resulting in the formation of a simplicial three-complex. Emergent properties of geometry and gravity arise due to this transition.

The model can be described by the Hamiltonian, which is a sum of two parts: $H_\tau = H_0 + H_s$. The first term, $H_0$, represents the Hamiltonian of the Fermi liquid of preons, and the second one,  $H_s$, describes the spacetime network. The Fermi liquid obeys the Fermi-Dirac distribution:
\begin{align}
    \bar n_{i }=\frac{1}{e^{\beta \left(\varepsilon_{i }-\mu_q\right) } +1},
    \label{FDQ}
\end{align}
where $\mu_q$ is the chemical potential, $\bar n_i$ refers to the average occupation number with the normalization $N = \sum_i \bar n_i$ and $N$ is the total number of particles, representing 0-simplices of the network. The Hamiltonian, $H_0$, can be written as $H_0 = - m_p \sum_i \bar n_i +\sum_i \varepsilon_i n_i$, where $n_i$ denotes the occupation state at site $i$,  The first term in $H_0$ accounts for the zero-point energy contribution. 

A simplicial complex allows for only one link between any two given nodes. Therefore, space can be represented only by fermionic graphs, where only one edge is permitted between any pair of vertices. To describe the spacetime network, we consider the modified version of the graph Hamiltonian proposed in \cite{NAHM}
\begin{align}\label{eq:H}
	H_s=   m_p \sum_{\langle ij \rangle} a_{ij}  - m_p \sum_{ \langle ijk \rangle} {\lambda}_{ijk} a_{ij}a_{jk}a_{ki} + m_p  \sum_{ \langle ijkp \rangle}  a_{ij}a_{jk}a_{kp}a_{pi} .
\end{align}
 Here the summation is over 1- simplices, 2-simplices, and 3-simplices denoted as $\langle ij \rangle$, $\langle ijk \rangle$ and $ \langle ijkp \rangle$, respectively. In \eqref{eq:H}, ${\lambda}_{ijk} = 1+  \epsilon_{ \langle ijk \rangle}\Delta S_{ \langle ijk \rangle}/R_{ \langle ijk \rangle}^2$ denotes the elementary holonomy of 2-simplex associated with the triplet of sites $\langle i,j,k \rangle$ forming a left triangle concerning the edge $\langle ij \rangle$, and  $\epsilon_{ \langle ijk \rangle}= 0,\pm 1$. 
 
 The first term in Eq.\eqref{eq:H} describes the contribution of the bosonic degree of freedom related to the two-atom entangled states, the second term yields the contribution of the quantum fermionic vacuum fluctuations and curvature of the entangled three-atom states, and the third one represents the quantum bosonic degree of freedom related to the entangled four-atom states. 

Within the mean-field approximation, the graph Hamiltonian $H_s$ is
replaced by the effective Hamiltonian
\begin{align}
 {\mathcal H}_s=   \sum_{ \langle ij\rangle} \varepsilon_{ij}a_{ij} +{2 m_p}\sum_{ \langle ijk \rangle} {\lambda}_{ijk}p_{ij} p_{jk} p_{ki} - 3{m_p}\sum_{ \langle ijkp \rangle} p_{ij} p_{jk} p_{kp} p_{pi}  ,
\label{H2}
\end{align}
where 
\begin{align}\label{EqEF}
	\varepsilon_{ij}= m_p- {3 m_p}\sum_{ k} \lambda_{ijk}p_{ik} p_{kj}+ {4 m_p}\sum_{ k,p} p_{ik}p_{kp} p_{pj} . 
\end{align}
The network is described by the generalized Fermi-Dirac distribution \eqref{FDBE} that can be recast as follows: 
\begin{align} \label{EqP1}
	p_{ij} = \frac{1}{2} \bigg( 1+ \tanh \bigg (\frac{\beta}{2} \Big ( \mu - m_p +  {3 m_p}\sum_{ k} \lambda_{ijk} p_{ik} p_{kj} -{4 m_p}\sum_{ k,p} p_{ik}p_{kp} p_{pj} \Big )\bigg ) \bigg)	
\end{align}  

At the Planck temperature, $T_p$, the system experiences the phase transition, leading to fundamental structural changes in the network topology \cite{NAHM}. The entanglement of the system is significantly reduced at temperatures above $T_p$. It is completely lost for the high temperatures ($T \gg T_p$), so the graph becomes disconnected, and a simpilicial 0-complex presents the space. Below the Planck temperature, the entanglement increases significantly, peaking at the system's ground state. Thus, the high dimensional simplicial complices emerge with decreasing the network's temperature, and the maximally triangulated space appears at zero temperature. For $T \gg T_p$, the ground state is defined by the Fermi liquid's ground state; for $T \ll T_p$, the system is in the bosonic ground state. Finally, it's worth noting that on a lattice, the motion of the preons is restricted, and their energy spectrum becomes discretized.
   \begin{figure}[tbh]
      	\begin{center}
      		\includegraphics[width=8cm]{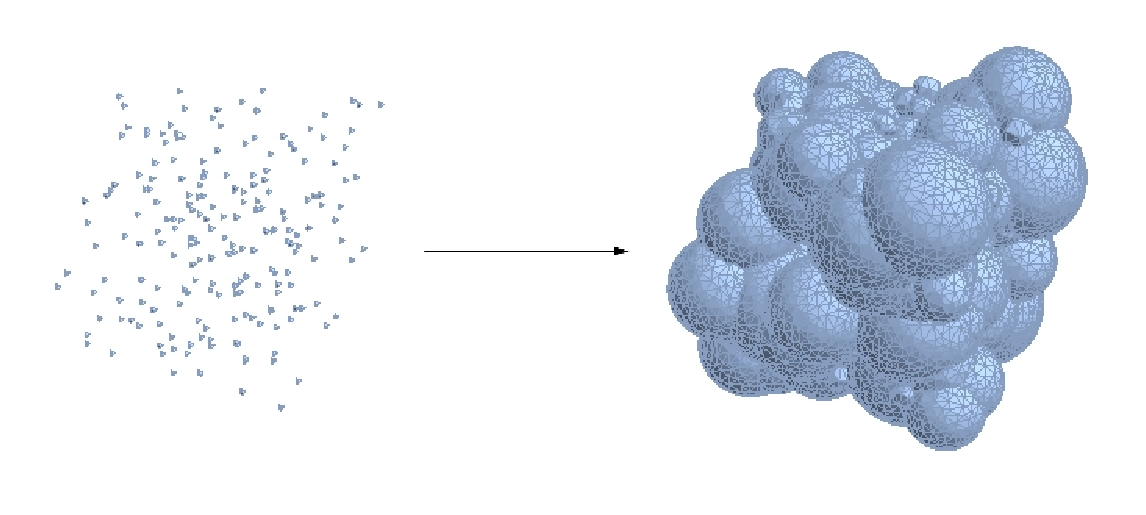}
      	\end{center}
      	\caption{Formation of a cluster from 200 initial randomly distributed tetrahedrons. Adapted with permission from Ref. \cite{NAMH}.}\label{Fig5}
      \end{figure}
   
   The energy of the total system is given by 
   \begin{align}\label{eq:E}
  E=   -N m_p +\sum_i \varepsilon_i \bar n_i + m_p \sum_{\langle ij \rangle} p_{ij}  - m_p \sum_{ \langle ijk \rangle} {\lambda}_{ijk} p_{ij}p_{jk}p_{ki} + m_p  \sum_{ \langle ijkp \rangle}  p_{ij}p_{jk}p_{kp}p_{pi} .
   \end{align}   

In the limit of $T \rightarrow 0$, we obtain the ground state with the energy
\begin{align}
E_{vac}= -  m_p \chi (\mathfrak M)-  m_p N_2 \frac{ \Delta \theta }{a^2 } ,
\label{EqE1}
\end{align}
where $\chi (\mathfrak M) = \sum^3_{i=0} (-1)^{i}  N_i -$  is the Euler characteristic of the simplicial complex $\mathfrak M$ with $N_i$ being a number of $i$-simplices \cite{KLST}, and $\Delta \theta$ is the average angle deficit of 2-simplices:
\begin{align}
	\Delta \theta= \frac{a^2}{N_2}\sum_{ \langle ijk \rangle} \epsilon_{ \langle ijk \rangle}\frac{\Delta S_{ \langle ijk \rangle}}{R_{ \langle ijk \rangle}^2}.
\end{align}
Here $a$ is the cosmic scale factor.

To find the effective pressure in our model, we use the relation $P = - \partial F/\partial V|_{T}$, where $F =E - TS $ is the Helmholtz free energy. In the limit of $T \rightarrow 0$, we obtain $P =-\rho_{vac}$, where $\rho_{vac} = E_{vac}/V$ is the density of vacuum energy. Thus, the pressure of the vacuum in the network is negative and should lead to the universe's accelerating expansion. 

{\em Evolution of the universe.} -- In our approach, space and time are quantized in the Planck units, and a random/stochastic process governs spacetime evolution. A random walk with a reflecting barrier $R_0$ at $n=0$ and a step $\ell_0$ defines the cosmological scale factor. After $n$ steps, the average cosmological scale factor, $\left\langle a_n\right\rangle$, grows as follows \cite{NAMH}
     \begin{align}
     	  \left\langle a_n\right\rangle=a_0+\sqrt{2 n / \pi} ,
     \end{align}
 where $a_0 = R_0/\ell_0 $ is the scale factor related to the initial 3-simplex defining the minimal size of the universe. After $\sim 10^8$ steps in Planck units of time, which corresponds to pre-inflationary time (time of the beginning of inflation), $t_{pr} \sim 10^{-35} \mathrm{sec}$ the cosmological scale factor would be $a_{p r} \sim a_0 + \sqrt{2/\pi}10^4$. 
 
One can expect the formation of clusters of various dimensions in the process of evolution. It is reasonable to assume that spacetime consists of many small fluctuating simplicial regions. These fluctuations may result in forming mini-clusters, as illustrated in Fig. \ref{Fig5}. They can be created in different areas and grow as mini-universes, forming spatial foam. 
  
\begin{figure}[tbh]
  	\centering
  	\includegraphics[width=0.7\linewidth]{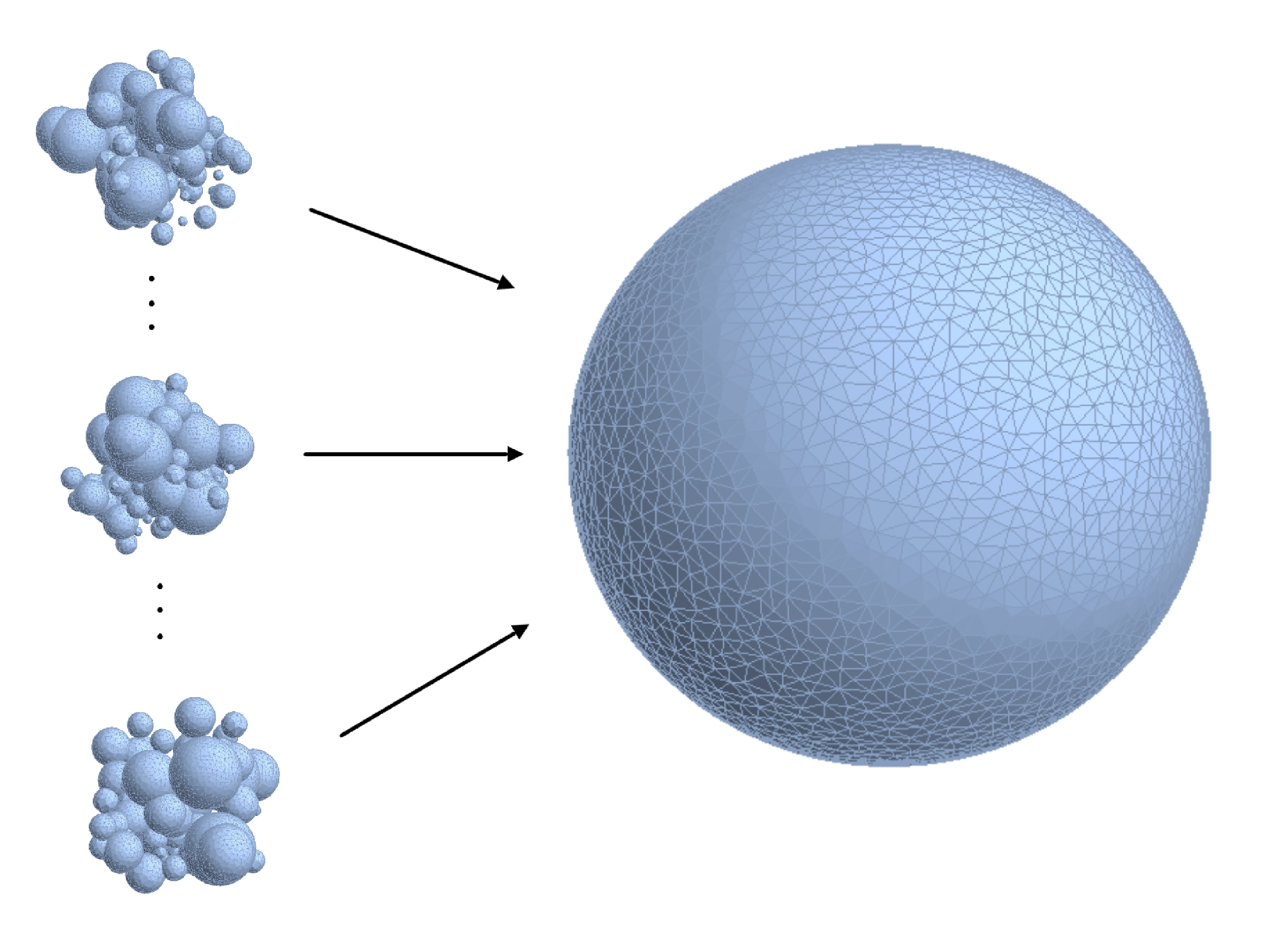}
  	\caption{Reconstruction of the Universe after inflation. Cosmic inflation predicts many big bangs, creating multiple expanding universes being part of a multiverse. Adapted with permission from Ref. \cite{NAMH}.}
  	\label{U}
  \end{figure}

   Thus, for $T < T_p$,  the graph is represented by a simplicial 3-complex formed by groups or clusters of various dimensions. Within the same group, node-node connections are dense, but connections are less dense between groups. A group of nodes forms a {\em component} when all of them are connected, directly or indirectly \cite{NMSW,LI20211}. A ``giant component" contains a significant part of the total number of nodes. For instance, if the pure simplicial complex presents the spacetime network, the whole network is a giant component. The current universe's formation occurs in the inflation period as a merging of the mini-universes and the formation of a giant component as the result of the structural phase transition occurring at the temperature $T_c \sim 10^{28}\rm K$ (Fig. \ref{U}). 
 
{\em The cosmological constant problem.} -- In classical general relativity, the cosmological constant, $\Lambda$,  added in the Einstein equations, governs the Universe's accelerated expansion. Writing the Einstein equations with the cosmological constant as
\begin{align}
	R_{\mu \nu }-\frac{1}{2} g_{\mu \nu } R={8 \pi G}T_{\mu \nu }-g_{\mu \nu } \Lambda.
	\label{EqL1}
\end{align}
one can observe that space with non-zero cosmological constant produces the same gravitational field as the matter with mass density $\rho_{\Lambda}=\Lambda / 8 \pi G$ and pressure  $P_{\Lambda}=-\rho_{\Lambda}$. Thus, one can speak about the energy density of the vacuum and its pressure. 

According to the equivalence principle of general relativity, the energy of the quantum vacuum fluctuations must produce a gravitational field, and the semiclassical Einstein equations describe their contribution: 
\begin{align}
	R_{\mu \nu }-\frac{1}{2} g_{\mu \nu } R={8 \pi G} \langle T_{\mu \nu }  \rangle-g_{\mu \nu } \Lambda,
	\label{EqLB}
\end{align}
where $\langle T_{\mu \nu }  \rangle =\langle 0|T_{\mu \nu}|0\rangle$ is the expectation value of the quantum vacuum energy-momentum tensor. Lorentz invariance requires that $\langle T_{\mu \nu }  \rangle$ takes the form
$\langle T_{\mu \nu }  \rangle = - \rho_{vac} g_{\mu \nu}$
where  $\rho_{vac}= \langle 0|T_{00}|0\rangle$ is the expectation value of the energy density of the matter fields in the vacuum state.  Employing Eq.\eqref{EqLB}, one can recast Eq.\eqref{EqL1} as
\begin{align}
	R_{\mu \nu }-\frac{1}{2} g_{\mu \nu } R={8 \pi G}  T_{\mu \nu }  -g_{\mu \nu } \Lambda_{\mathrm{eff}} ,
	\label{EqLA}
\end{align}
where $\Lambda_{\mathrm{eff}} = \Lambda+8 \pi \rho_{vac} \ell_p^2$ is the effective (observable) cosmological constant. 

The current Hubble expansion rate $H_0$ yields an upper bound for $\Lambda_{\mathrm{eff}}$ \cite{WAQI},
\begin{align}
	\Lambda_{\mathrm{eff}}=3 H^2 \leq 3 H_0^2 \sim \Lambda_0 \ell_p^{-2},
	\label{EqLC}
\end{align}
where  $\Lambda_0 =10^{-122}$ is the dimensionless cosmological constant,  and $H =\dot a/a= \sqrt{\Lambda_{\mathrm{eff}}/3}$ is the Hubble constant describing acceleration of the universe. The vacuum density energy in the QFT is estimated as $\rho_{vac} \sim m_p^4$, where $m_p$ is the Planck mass. Using this result in \eqref{EqLC}, we obtain $\Lambda_{\mathrm{eff}} \sim  10^{-122} \rho_{vac} \ell_p^2$. Thus, one has a huge disagreement between the observed value of the cosmological constant and the theoretical prediction of its value resulting from the quantum field theory. 

It is worth noting that the standard computation of the cosmological constant assumes that the spacetime is homogeneous and isotropic, and the theory is Lorentz invariant. These assumptions are reasonable at a cosmological scale but questionable at a small (Planck) scale \cite{WQZZ,WAQI,CARS}. 

To resolve the cosmological constant problem, we use the statistical description of the spacetime, treating the latter as a complex network. We define the observable (effective) cosmological constant as
\begin{align}
	\Lambda = 8 \pi \ell_p^2\rho_{vac} \equiv \Lambda_0 \ell_p^{-2},  
	\label{EqL}
\end{align}
where $\rho_{vac} $ is the energy density of the vacuum state defined by the contribution of all bosonic and fermionic fields of the simplicial manifold $\mathfrak M$.

Employing Eq. \eqref{EqE1}, we obtain
\begin{align}
\rho_{vac} =  m_p n_g - \frac{m_p N_2}{V} \frac{ \Delta \theta }{a^2 }  ,
\label{EqR1}
\end{align}
where $n_g$ denotes the number density of topological geons:
\begin{align}
	n_{g}  =& - \frac{1}{V} \chi (\mathfrak M) 
	\label{EqG}
\end{align}
Substituting $V=N_3 V_0$ in Eq. \eqref{EqR1}, where $V_0 ={\ell_p^3}/{6 \sqrt{2}}$ is an average volume of the 3-simplex, and employing Eq.\eqref{EqL}, we obtain
\begin{align}
\Lambda_0 = 8 \pi\ell^3_pn_g
-  \frac{ 8\pi \ell^3_p N_2}{V_0 N_3}\cdot\frac{ \Delta \theta }{a^2}.
\label{EqL2a}
\end{align}

Assume that  $\mathfrak M$ is a pure 3-dimensional simplicial complex, then $\chi (\mathfrak M) =0$ and $N_2 = 2 N_3$ \cite{KLST}. Using these results, we obtain 
\begin{align}
\Lambda_0 = -  \frac{ 16\pi \ell^3_p }{V_0 }\cdot\frac{ \Delta \theta }{a^2}.
\label{EqL3}
\end{align}
Thus, if the simplicial manifold is represented by a pure 3-dimensional simplicial complex before inflation, the cosmological constant is determined by its average curvature. In another limit case, when the contribution of the second term in Eq. \eqref{EqL2a} is negligible in comparison with the first term, we obtain $\Lambda_0 = {8 \pi\ell^3_p}n_g$. The possible scenario of the cosmological constant emerging is as follows. Before inflation, the cosmological constant is determined by the average curvature of the universe. During the inflation period, the term with $\chi(\frak M)$ begins to dominate so that at the end of the inflation, the cosmological constant is determined by the density of topological geons.

The density of the topological geons determines the effective cosmological constant. The current observables yield $n_g \approx 10^{-25}\rm cm^{-3}$. Taking the size of the observable universe as 93 billion light-years, we find that the number of the geons in our universe, defined by the Euler characteristic, is $\sim 10^{61}$. This enormous number implies that the topology of spacetime is highly non-trivial due to vacuum fluctuations. 

In summary, we point out that the standard formulation of the cosmological constant problem is problematic since spacetime and vacuum energy density are highly inhomogeneous and fluctuate wildly at the Planck scale. The vacuum fluctuations can result in the foamy structure of spacetime, formed by micro-universes connected by the  Einstein-Rosen bridges with wormhole topology. We show that the spacetime foam significantly contributes to the effective cosmological constant defined by the density of topological geons. Our research reveals the role of spacetime topology in solving the cosmological constant problem and demonstrates that the source of the dark energy is topological geons.

The proposed model challenges the current Quantum Field Theory approach by offering a statistical description of spacetime and suggests a new approach to address the cosmological constant problem. The implications of this statistical description of spacetime are significant and could reshape our understanding of the universe's fundamental nature. \\

The author acknowledges the support of the CONAHCYT.

%


 \end{document}